\documentclass[conference]{IEEEtran}

\IEEEoverridecommandlockouts
\usepackage{cite}
\usepackage{amsmath,amssymb,amsfonts}
\usepackage{textcomp}
\usepackage{xcolor}
\def\BibTeX{{\rm B\kern-.05em{\sc i\kern-.025em b}\kern-.08em
    T\kern-.1667em\lower.7ex\hbox{E}\kern-.125emX}}

\usepackage{epsfig}
\usepackage{color}
\usepackage{amsmath} 
\usepackage{amssymb}
\usepackage{amsxtra}
\usepackage{enumitem}

\usepackage[T1]{fontenc}
\usepackage{setspace}

\usepackage{amsmath}
\usepackage[cmintegrals]{newtxmath}
\usepackage{graphicx}
\graphicspath{{Figures/}}
\usepackage{url}
\usepackage{cite}
\usepackage{algorithm}
\usepackage{algpseudocode}

\makeatletter
\def\BState{\State\hskip-\ALG@thistlm}
\makeatother

\usepackage{epstopdf}
\usepackage{amsmath}
\usepackage{amsxtra}
\usepackage{amstext}
\usepackage{amssymb}
\usepackage{latexsym}
\usepackage{dsfont} 
\usepackage{color}
\usepackage{multirow}
\usepackage{float}
\usepackage{hyperref}

\usepackage{tabularx,colortbl}

\usepackage{lscape}

\usepackage{algorithm}
\usepackage{algpseudocode}

\usepackage{units}
\usepackage{cases}
\usepackage[ subrefformat=parens,labelformat=parens, caption=false, font=footnotesize]{subfig}
\usepackage{mathtools}

\newcommand{\mbf}[1]{\mathbf{#1}}

\newcommand{\nth}[1]{{#1}{\text{th}}}

\newcommand{\abs}[1]{\left|{#1}\right|}
\newcommand{\norm}[1]{\left\|{#1}\right\|}

\DeclareMathOperator*{\argmax}{arg\,max}   
\DeclareMathOperator*{\argmin}{arg\,min}

\usepackage{mathtools}

\newcommand{\ML}{\mathrm{ML}}
\newcommand{\ZF}{\mathrm{ZF}}

\newcommand{\MAP}{\mathrm{MAP}}
\newcommand{\cML}{\mathrm{cML}}

\newcommand{\GRAND}{\mathrm{GRAND}}

\newcommand{\Prob}{\mathrm{Pr}}

\newcommand{\CMT}{\mathrm{CMT}}
\newcommand{\CAD}{\mathrm{CAD}}

\newfont{\bb}{msbm10 scaled 1100}
\newcommand{\CC}{\mbox{\bb C}}

\newcommand{\RR}{\mbox{\bb R}}

\newcommand{\FF}{\mbox{\bb F}}

\begin{document}

\title{Soft-input, soft-output joint detection and GRAND}

\author{\IEEEauthorblockN{Hadi~Sarieddeen}
\IEEEauthorblockA{\textit{Research Laboratory of Electronics} \\
\textit{Massachusetts Institute of Technology}\\
Cambridge, MA 02139, USA \\
hadisari@mit.edu}
\and
\IEEEauthorblockN{Muriel M{\'e}dard}
\IEEEauthorblockA{\textit{Research Laboratory of Electronics} \\
\textit{Massachusetts Institute of Technology}\\
Cambridge, MA 02139, USA \\
medard@mit.edu}
\and
\IEEEauthorblockN{Ken. R. Duffy}
\IEEEauthorblockA{\textit{Hamilton Institute} \\
\textit{Maynooth University}\\
Ireland \\
ken.duffy@mu.ie}
\thanks{The project or effort depicted was or is sponsored by the Defense Advanced Research Projects Agency under Grant number HR00112120008. The content of the information does not necessarily reflect the position or policy of the Government, and no official endorsement should be inferred.

To appear in the IEEE GLOBECOM 2022 proceedings.}
}


\maketitle

\begin{abstract}

Guessing random additive noise decoding (GRAND) is a maximum likelihood (ML) decoding method that identifies the noise effects corrupting code-words of arbitrary code-books. In a joint detection and decoding framework, this work demonstrates how GRAND can leverage crude soft information in received symbols and channel state information to generate, through guesswork, soft bit reliability outputs in log-likelihood ratios (LLRs). The LLRs are generated via successive computations of Euclidean-distance metrics corresponding to candidate noise-recovered words. Noting that the entropy of noise is much smaller than that of information bits, a small number of noise effect guesses generally suffices to hit a code-word, which allows generating LLRs for critical bits; LLR saturation is applied to the remaining bits. In an iterative (turbo) mode, the generated LLRs at a given soft-input, soft-output GRAND iteration serve as enhanced a priori information that adapts noise-sequence guess ordering in a subsequent iteration. Simulations demonstrate that a few turbo-GRAND iterations match the performance of ML-detection-based soft-GRAND in both AWGN and Rayleigh fading channels at a complexity cost that, on average, grows linearly (instead of exponentially) with the number of symbols.

\end{abstract}


\section{Introduction}
\label{sec:introduction}

The upcoming sixth generation (6G) of wireless communications promises to support a plethora of data-demanding and delay-sensitive applications \cite{rajatheva2020white}, requiring both ultra-broadband high-frequency connectivity \cite{Sarieddeen9514889} and ultra-reliable low-latency communication (URLLC) \cite{Durisi7529226}. Such variety in requirements compels a paradigm shift from structured, code-specific channel-code decoding to universal and practical decoding that is efficient for different code rates and lengths. Early universal near-maximum-likelihood (ML) decoders for linear codes adopted a list-decoding principle \cite{Gazelle567691,Valembois1291728}. Recently, guessing random additive noise decoding (GRAND) \cite{Duffy8437648,Duffy8630851} is celebrated as a novel and practical universal decoder suited for block-code constructions of moderate redundancy. 


GRAND is a capacity-achieving channel-code decoder that has demonstrated ML decoding performance on arbitrary (even unstructured) code-books. Instead of directly identifying the transmitted code-word, GRAND aims at identifying the noise effect that corrupts the code-word; it successively reverses the noise effects from the received signal to recover candidate transmitted words. By leveraging information on channel and noise models, the candidate noise sequences are ordered and queried in decreasing likelihood, guaranteeing the first recovered code-word to be the ML decoding solution, even for channels with memory in the absence of interleaving. The guesswork literature \cite{Christiansen6340341,Beirami8522043,Arikan481781} establishes the computational feasibility of GRAND for all moderate redundancy codes, where the Shannon entropy rate of noise is typically less than that of information symbols \cite{Duffy8630851}. Furthermore, GRAND's computational efficiency and modularity have resulted in highly efficient circuit design, as demonstrated in a recent 65 nm \cite{Abbas9195254} synthesis and a 40 nm \cite{Riaz9567867} CMOS implementation.

Incorporating soft-detection symbol reliability information into decoding decisions enhances decoding accuracy \cite{Kaneko605601,Fossorier412683}. At one end, GRAND can leverage as soft information a one-bit mask designating whether a channel use is reliable or not \cite{Duffy8849297}; specifying reliable bits via a channel-fading-induced mask is also proposed in \cite{abbas2022grand}. At the other end, the complete information in continuous channel outputs serves as soft symbol-reliability information in the soft-GRAND (SGRAND) scheme \cite{Solomon9149208}. A compromise between one-bit soft-GRAND and SGRAND is ordered reliability bits GRAND (ORBGRAND) \cite{Duffy9414615}, which matches the decoding accuracy of SGRAND through code-book-independent quantization of soft information in a hardware-friendly algorithm. 

With fading channels, generating high-resolution soft information (as opposed to masks \cite{abbas2022grand}) through exhaustive ML detection is computationally demanding, and low-complexity soft-output detectors only generate sub-optimal LLRs. In the absence of soft information (or with low-quality information), iterative decoding schemes can intrinsically generate soft-decoding reliability information to be fed as soft-input decoding information in subsequent iterations \cite{Berrou539767}. Such information can be updated over both the detection and decoding iterations \cite{Tomasoni5581209}, introducing more degrees of freedom in versatile, adaptive communication systems. However, how to generate soft-reliability outputs through GRAND remains unclear. Consequently, iterative soft-input, soft-output (SISO) GRAND (turbo-GRAND) has not yet been investigated.


In this work, we propose a variation of GRAND that does not avail of input soft-detection bit reliability information but leverages complex received symbols (soft-information in a crude, unprocessed form), channel state information (CSI), and demapped bits (linear detector outputs) to generate soft-decoding bit-reliability information in log-likelihood ratios (LLRs). We compute the LLRs by populating Euclidean distance metrics corresponding to a list of candidate words--not necessarily code-words--resulting from noise guesswork. Because the number of guesses before hitting a code-word is limited, we cannot compute LLRs for all bits; we propose LLR thresholding. We propose a turbo-GRAND scheme in which the generated extrinsic LLRs are fed as a priori LLRs in subsequent SISO decoding iterations. With access to complex received vectors and continuous CSI, the Euclidean distance computations are common to both detection and decoding; turbo-GRAND thus realizes joint detection and decoding.

The paper is organized as follows. The problem formulation is first presented in Sec.~\ref{sec:sysmodel}. Then, the proposed turbo-GRAND scheme is detailed in Sec.~\ref{sec:proposed}. Performance and complexity results are reported in Sec.~\ref{sec:results}; conclusions are drawn in Sec. \ref{sec:conc}. Regarding notation, bold upper case, bold lower case, and lower case letters correspond to matrices, vectors, and scalars, respectively. Scalar and vector $\text{L}_2$ norms are denoted by $\abs{\cdot}$ and $\norm{\cdot}$, respectively. $\mathsf{E}[\cdot]$, $(\cdot)^{T}$, and $(\cdot)^{*}$, stand for the expected value, transpose, and conjugate transpose, respectively. $\mbf{I}_M$ is an identity matrix of size $M$, $\mbf{0}_N$ is a vector of zeros of size $N$, $\FF_u$ denotes a Galois field with $u$ elements, $\Prob(\cdot)$ is the probability function, and $\odot$ is the Hadamard product.

\section{System Model and Problem Formulation}
\label{sec:sysmodel}

\subsection{System Model}
\label{sec:system}

We consider a communication system of equivalent baseband input-output relation, $\mbf{y} = \mbf{H}\mbf{x} + {\mbf{n}}$,
where $\mbf{y}\!\in\!\CC^{M\times1}$ is the received symbol vector, $\mbf{H}\!\in\!\CC^{M\times M}$ is the channel matrix, $\mbf{x}\!=\![x_{1}\cdots x_{i}\cdots x_{M}^{}]^{T}\!\in\!\CC^{M\times1}$ is the transmitted symbol vector, and ${\mbf{n}}\!=\![{n}_{1}\cdots {n}_{i}\cdots {n}_{M}^{}]^{T}\!\in\!\CC^{M\times1}$ is the additive--possibly colored--noise vector $\left(\mathsf{E}[{n}_i {n}_i^{*}]\!=\!\sigma_i^{2}\right)$. 
Note that $\mbf{H}$ can be an identity matrix in an additive white Gaussian noise (AWGN) system, a diagonal matrix under point-to-point fading, or a complete matrix under spatial diversity/multiplexing schemes.
Furthermore, we assume the information symbol, $x_{i}$, to belong to a normalized complex constellation, $\mathcal{X}_{i}$ $\left(\mathsf{E}[x_{i}^{*}x_{i}^{}]\!=\!1\right)$. Consequently, $\mbf{x}\!\in\! \bar{\mathcal{X}} \subset \CC^{M\times1}$, where $\bar{\mathcal{X}}$ is the finite $M$-dimensional lattice of all possible modulated symbol vectors.
The bit-representation of $x_i$ is $\mbf{c}_{i}\!=\![c_{i,1}\cdots c_{i,j}\cdots c_{i,q_i}]^{T}\!\in\! \FF_2^{q_i}$, where $q_i\!=\!\lceil{\log_2(\abs{\mathcal{X}_i})\rceil}$. The bit-representation of $\mbf{x}$ is thus $\mbf{c}\!=\![\mbf{c}_{1}\cdots \mbf{c}_{i} \cdots \mbf{c}_{M}]^{T}\!\in\!\FF_2^N$, where $N\!=\! \sum_{i=1}^{M} q_i$. We assume $\mbf{c}$ to be a code-word encoded with an error correcting code $\alpha\!:\! \FF_2^K \!\rightarrow\! \FF_2^{N}$, of code-rate $R\!=\!K/N$. A code-book $\mathcal{C}\!\triangleq\!\{ \mbf{c}\!: \mbf{c} \!=\! \alpha(\mbf{b}), \mbf{b}\!\in\!\FF_2^K\}$ includes all possible code-words, where $\mbf{b}$ is the uncoded bit vector. We denote by $\mbf{v}\!\in\!{\RR^+}^N$ a vector of $\sigma_{i,j}^2$ values, the second-order noise statistics per bit. At the receiver side, assuming perfect CSI, a hard detector, $\beta:\CC^M\!\rightarrow\!\bar{\mathcal{X}}$, equalizes the channel and recovers a symbol vector, $\hat{\mbf{x}}$, from $\mbf{y}$; a demapper recovers a word, $\hat{\mbf{c}}$, from $\hat{\mbf{x}}$.

\subsection{Problem Formulation}
\label{sec:problem}

\begin{algorithm}[t]
\caption{Hard GRAND}\label{alg:GRAND}
\begin{algorithmic}[1]

\Require Demapped bits $\hat{\mbf{c}}$; ordered noise-generating function $\Pi$; abandonment threshold $B$
\Ensure Decoded $\bar{\mbf{c}}^{\GRAND}$
\State $k \gets 0$
\While{$k < 2^N$}
\State $k \gets k+1$
\State $\mbf{w} \gets \Pi(k)$ \Comment{ $\nth{k}$ likely noise sequence}
\If{$\hat{\mbf{c}}\ominus \mbf{w} \in \mathcal{C}$ \textbf{or} $k=B$}
    \State $\bar{\mbf{c}}^{\GRAND} \gets \hat{\mbf{c}}\ominus \mbf{w}$
    \State \textbf{return} $\bar{\mbf{c}}^{\GRAND}$
\EndIf
\EndWhile

\end{algorithmic}
\end{algorithm}

The ML decoder computes the conditional probability of the demapped word, $\hat{\mbf{c}}$, for each of the $2^K$ code-words, $\mbf{c}$, in code-book, $\mathcal{C}$. The $\mbf{c}$ with the highest conditional likelihood of transmission given what was received is the ML solution, $\bar{\mbf{c}}^{\ML} \!=\! \argmax \{ \Prob \left( \hat{\mbf{c}}\mid \mbf{c}\right)\!:\! \mbf{c} \!\in\! \mathcal{C}\}$.
Instead of searching code-words, GRAND searches putative, not necessarily memoryless, noise effect sequences that corrupt $\mbf{c}$, $\mbf{w}\!=\![w_{1}\cdots w_{i,j}\cdots w_{N}^{}]^{T}\!\in\!\FF_2^N$, with non-increasing probability. We express the channel's action at the bit level through function $\oplus$, where $\bar{\mbf{c}} = \mbf{c}\oplus \mbf{w}$. We can write $\Prob \left(\bar{\mbf{c}}\!\mid\! \mbf{c}\right) = \Prob \left(\bar{\mbf{c}} =  \mbf{c}\oplus \mbf{w}\right)$, and it follows that
\begin{equation}\label{eq:GRAND}
\bar{\mbf{c}}^{\GRAND} = \argmax \{ \Prob \left(\mbf{w}= \bar{\mbf{c}} \ominus \mbf{c}\right): \mbf{c} \in \mathcal{C}\}.
\end{equation}
The receiver creates a list of noise effect sequences of decreasing order of likelihood through a noise generating function $\Pi\!:\! \{1,\! \cdots\!, 2^N\} \!\rightarrow\! \mbf{w}\!\in\! \FF_2^N\!$; the sequences are queried until the first code-word hit, $\mbf{w}\!=\! \hat{\mbf{c}} \!\ominus\! \mbf{c}$ (block-code syndrome computations). GRAND is thus a ML decoder that executes Algorithm~\ref{alg:GRAND} and returns $\bar{\mbf{c}}^{\GRAND}$; information bits are retrieved as $\bar{\mbf{b}}^{\GRAND} \!=\! \alpha^{-1}(\bar{\mbf{c}}^{\GRAND})$. Because the entropy of noise is small in most communication systems, GRAND is low-complexity. GRAND's efficiency is further guaranteed by abandoning guessing after a computational cut-off \cite{Duffy8630851}, $B$.  

Soft GRAND accepts, in addition to $\hat{\mbf{c}}$, bit-reliability information in a vector $\mbf{\Lambda}\!=\![\lambda_{1,1}\cdots \lambda_{i,j}\cdots \lambda_{M,q}^{}]^{T}\!\in\!\RR^N$ (assuming $q_i\!=\!q, \forall i$). We can generate a weight metric per putative noise sequence by multiplying the noise sequences by $\abs{\mbf{\Lambda}}$; noise sequences with smaller weights are more likely to occur. However, $\mbf{\Lambda}$ is not always available (or available but with bad quality), as soft-output detectors are computationally demanding. We aim at generating soft-reliability outputs within GRAND, proposing a low-complexity noise-centric soft-output decoding algorithm, a function $\bar{\alpha}: \{\FF_2^K,\CC^{M}\} \!\rightarrow\! \RR^{N}$, that accepts $\hat{\mbf{c}}$ and $\mbf{y}$ and generates output LLRs, $\bar{\mbf{\Lambda}}\!=\![\bar{\lambda}_{1,1}\cdots \bar{\lambda}_{i,j}\cdots \bar{\lambda}_{M,q}^{}]^{T}\!\in\!\RR^N$, with the knowledge of $\mbf{H}$. By further integrating knowledge of noise statistics in LLR computations (scaling by noise variance), our proposal can account for noise bursts (Markovian channel noise, for example), foregoing interleaves, and whitening filters \cite{Duffy8630851,An9322303}. The extrinsic LLRs, $\bar{\mbf{\Lambda}}$, can then be fed as intrinsic LLRs, $\mbf{\Lambda}$, in a subsequent soft GRAND; an iteration that is repeated in the proposed turbo-GRAND.

\section{Proposed SISO Turbo-GRAND}
\label{sec:proposed}

\begin{figure}[!t]
  \centering
  \includegraphics[width=3.5in]{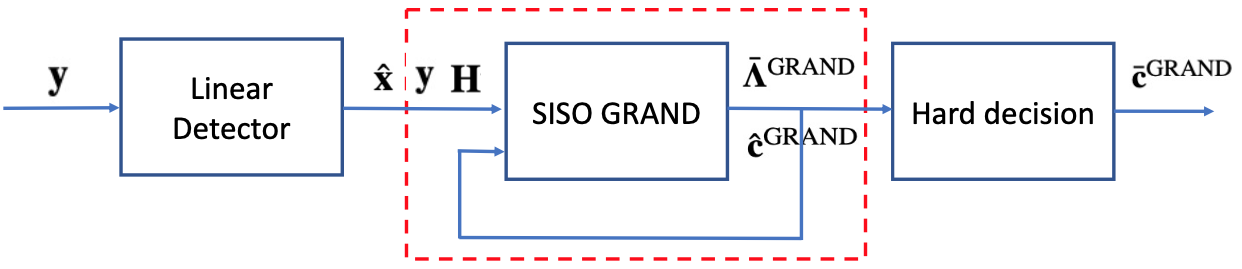}
  \caption{A block diagram of turbo-GRAND.}\label{fig:GRAND_turbo}
\end{figure}

We propose a SISO variation of GRAND that leverages, in addition to the resources of Algorithm \ref{alg:GRAND}, the received symbols, CSI, and optional noise statistics, to generate extrinsic bit-reliability LLRs through joint detection and decoding. The LLRs are fed as input soft-decoding information to a subsequent SISO GRAND iteration (up to $T$ iterations), resulting in turbo-GRAND (Fig.~\ref{fig:GRAND_turbo}). Turbo-GRAND aims to approach the performance of a soft GRAND with soft-input information from an exhaustive soft-output ML detector. Therefore, turbo-GRAND is helpful in the absence of soft-input information or the presence of sub-optimal soft information.

The LLR of bit $j$ of symbol $i$ is defined as
\begin{equation}\label{eq:LLR_ML1}
    \lambda_{i,j} = \log \left(\Prob\left( c_{i,j} \!=\! 1, \mbf{y},\mbf{H}\right)/\Prob\left( c_{i,j} \!=\! 0, \mbf{y},\mbf{H}\right)\right).
\end{equation}
Near-optimal ML-detection log-max LLRs \cite{Ivanov7436797} are computed by exhaustively searching the lattice $\bar{\mathcal{X}}$, computing $\abs{\mathcal{X}_1}\times\abs{\mathcal{X}_i}\times\cdots\times\abs{\mathcal{X}_M}$ Euclidean distance metrics to solve for \cite{8186206Sarieddeen}
\begin{equation}\label{eq:LLR_ML2}
  \lambda_{i,j}^{\ML} \approx \frac{1}{\sigma^{2}} \left(\min_{\mbf{x} \in \bar{\mathcal{X}}^{i,j,1}}{\norm{\mbf{y} - \mbf{H}\mbf{x}}^{2}} - \min_{\mbf{x} \in \bar{\mathcal{X}}^{i,j,0}}{\norm{\mbf{y} - \mbf{H}\mbf{x}}^{2}} \right),
\end{equation}
where $\bar{\mathcal{X}}^{i,j,1}\!\triangleq\!\{\mbf{x} \!\in\! \bar{\mathcal{X}}: c_{i,j}\!=\!1\}$ and $\bar{\mathcal{X}}^{i,j,0}\!\triangleq\!\{\mbf{x} \!\in\! \bar{\mathcal{X}}: c_{i,j}\!=\!0\}$ are subsets of symbol vectors in $\bar{\mathcal{X}}$, having in the corresponding $\nth{j}$ bit of the $\nth{i}$ symbol a value of $1$ and $0$, respectively. We have further assumed the case of white noise, $\sigma_{i,j}\!=\!\sigma, \forall i,j$. For colored noise, $\norm{\mbf{y} - \mbf{H}\mbf{x}}^{2}$ is replaced by $\left(\mbf{y} - \mbf{H}\mbf{x}\right)^*\Gamma^{-1}\left(\mbf{y} - \mbf{H}\mbf{x}\right)$, where $\Gamma \!=\! \text{diag}\left(\mbf{v}\right)$. The hard ML detection output is $\hat{\mbf{x}}^{\ML} \!=\! \argmin_{\mbf{x} \in \bar{\mathcal{X}}}\norm{\mbf{y} - \mbf{H}\mbf{x}}^{2}$.
Furthermore, soft information can be extracted per symbol in linear detectors \cite{studer2011asic}, which are near-optimal in point-to-point systems at a high signal-to-noise ratio (SNR) but sub-optimal under symbol interference/correlation. In particular, a zero-forcing (ZF) detector equalizes the channel by multiplying by its pseudo-inverse, $\mbf{\hat{y}}^{\ZF} \!=\! \left(\mbf{H}^{*}\mbf{H}\right)^{-1} \mbf{H}^{*}\mbf{y}$. The per-symbol ZF LLRs in $\mbf{\Lambda}^{\ZF}$ are
\begin{equation}\label{eq:LLR_ZF}
\lambda_{i,j}^{\ZF} = \frac{1}{{\sigma_{i}^{\ZF}}^2} \left(\min_{x_i\in \mathcal{X}_i^{j,1}} \abs{\hat{y}_i^{\ZF} - x_i}^{2} - \min_{x_i\in \mathcal{X}_i^{j,0}} \abs{\hat{y}_i^{\ZF} - x_i}^{2}\right),
\end{equation}
where $\mathcal{X}_i^{j,1}\!\triangleq\!\{x_i \!\in\! \mathcal{X}_i: c_{i,j}\!=\!1\}$ and $\mathcal{X}_i^{j,0}\!\triangleq\!\{x_i \!\in\! \mathcal{X}_i: c_{i,j}\!=\!0\}$ are subsets of symbols in the one-dimensional constellation $\mathcal{X}_i$, having a $\nth{j}$ bit of $1$ and $0$, respectively, and ${\sigma_{i}^{\ZF}}^2\!=\!\sigma_i^{2} \!\left(h_i^{*} h_i\right)^{-1}$ is a scaled noise variance. The ZF hard-outputs are computed per symbol as $\hat{x_i}^{\ZF} \!=\! \left\lfloor \hat{y}^{\ZF}_i- x_i \right\rceil_{\mathcal{X}_i}$, where $\lfloor \eta \rceil_{\mathcal{X}_i} \triangleq \argmin_{x \in \mathcal{X}_i} \abs{\eta-x}$ is the slicing operator over $\mathcal{X}_i$. The demapped version of $\hat{\mbf{x}}^{\ZF}$, $\hat{\mbf{c}}^{\ZF}$, is the initial vector over which noise effects are guessed in turbo-GRAND.

\begin{algorithm}[!t]
\caption{Joint detection and decoding - turbo-GRAND}\label{alg:turboGRAND}
\begin{algorithmic}[1]

\Require Demapped bits $\hat{\mbf{c}}\!=\!\hat{\mbf{c}}^{\ZF}$; received symbols $\mbf{y}$; channel matrix $\mbf{H}$; noise matrix $\mbf{W}$; noise statistics $\mbf{v}$; noise generating/sorting function $\Pi/\acute{\Pi}$; input LLRs $\mbf{\Lambda}$ ($\mbf{\Lambda} \!=\! \mbf{\Lambda}^{\ZF}$ or $\mbf{\Lambda} \!=\! \mbf{0}_N$); abandonment threshold $B$
\Ensure Output LLRs $\bar{\mbf{\Lambda}}^{\GRAND}$; demapped $\hat{\mbf{c}}^{\GRAND}$; decoded $\bar{\mbf{c}}^{\GRAND}$

\State $d^{\ML} \!\gets\! \infty$;  $\bar{d}^{\ML} \!\gets\! \infty$; $\mbf{d}^{\cML} \!\gets\! N\odot(1/\mbf{v})$
\State $\hat{\mbf{c}}^{\GRAND} \!\gets\! \hat{\mbf{c}}$; $\bar{\mbf{c}}^{\GRAND} \!\gets\! \hat{\mbf{c}}$; $\bar{\mbf{\Lambda}}^{\GRAND} \!\gets\! \mbf{\Lambda}$

\State $t \gets 0$;  
\While{$t < T$}

\State $\mbf{s} \gets \acute{\Pi}\left(\mbf{W}\times\abs{\bar{\mbf{\Lambda}}^{\GRAND}}\right)$ 
\State $k \gets 0$

\While{$k < B$}
\State $k \gets k+1$; $\mbf{w} \gets \Pi(\mbf{s}(k))$ \Comment{ $\nth{k}$ likely noise}
\State $\bar{\mbf{c}} \gets \hat{\mbf{c}}^{\GRAND}\ominus \mbf{w}$; $\bar{\mbf{x}} \gets \text{mod}\left(\bar{\mbf{c}}\right)$
\State $d \gets \left(\mbf{y} - \mbf{H}\bar{\mbf{x}}\right)^*\Gamma^{-1}\left(\mbf{y} - \mbf{H}\bar{\mbf{x}}\right)$ \Comment{$\Gamma \!=\! \text{diag}\left(\mbf{v}\right)$}

    \State $n \gets 0$
    \While{$n < N$}
            \State $n \gets n+1$
            \If{$\bar{c}_{n} \neq \hat{c}^{\GRAND}_{n}$}\Comment{complementary bits}
                \If{$d < d^{\ML}$}
                    \State $d^{\cML}_{n} \gets d^{\ML}$\Comment{update $d^{\cML}$ values}
                \ElsIf{$d < d^{\cML}_n$}
                    \State $d^{\cML}_{n} \gets d$
                \EndIf
            \EndIf
    \EndWhile
    \If{$d < d^{\ML}$} \Comment{detection}
        \State $\hat{\mbf{c}}^{\GRAND} \gets  \bar{\mbf{c}}$; $d^{\ML} \gets d$ 
    \EndIf

    \If{$\bar{\mbf{c}} \in \mathcal{C}$ \textbf{or} $k=B$}
    \Comment{code-word hit}
        \If{$d < \bar{d}^{\ML}$} \Comment{decoding}
            \State $\bar{\mbf{c}}^{\GRAND} \gets \bar{\mbf{c}}$; $\bar{d}^{\ML} \gets d$ 
        \EndIf
    \State $\bar{\mbf{\Lambda}}^{\GRAND} \gets \left(2\hat{\mbf{c}}^{\GRAND} - 1\right)\odot\left(d^{\ML} - \mbf{d}^{\cML}\right)$ 
    \State $t \gets t+1$
    \State \textbf{break} \Comment{go to next turbo iteration - line 4}
    \EndIf

    \EndWhile

\EndWhile
\State \textbf{return} $\bar{\mbf{\Lambda}}^{\GRAND}$, $\hat{\mbf{c}}^{\GRAND}$, \text{and} $\bar{\mbf{c}}^{\GRAND}$
\end{algorithmic}
\end{algorithm}



Contrary to conventional ML and list decoders that query all or multiple code-words, the basic implementation of GRAND achieves ML decoding performance by querying noise sequences and recovering a single code-word. This scarcity in code-word hits across turbo-GRAND iterations is mitigated in joint detection and decoding. We propose generating soft-output LLRs by populating a number of Euclidean distance computations equal to the number of noise guesses, as opposed to the exponential number of distance computations in \eqref{eq:LLR_ML2}. In particular, through guesswork, we aim at extracting an updated hard-detected vector, $\hat{\mbf{c}}^{\GRAND}$, and a reliability metric for each bit in $\hat{\mbf{c}}^{\GRAND}$, accumulated in the soft-decoding LLR vector, $\bar{\mbf{\Lambda}}^{\GRAND}$, all while recovering the decoded output, $\bar{\mbf{c}}^{\GRAND}$. 

Algorithm \ref{alg:turboGRAND} illustrates turbo-GRAND in more detail. For detection, we keep track of an ML-detection distance metric, $d^{\ML}$; for decoding, we keep track of an ML-decoding distance metric, $\bar{d}^{\ML}$. After each turbo-GRAND iteration, the candidate vector corresponding to $d^{\ML}$ is the updated detected vector, $\hat{\mbf{c}}^{\GRAND}$, and the candidate vector corresponding to $\bar{d}^{\ML}$ is the updated decoded vector, $\bar{\mbf{c}}^{\GRAND}$. Furthermore, for LLR computations, we accumulate a vector of counter-ML-detection distance metrics, $\mbf{d}^{\cML}\!=\![d^{\cML}_{1,1}\cdots d^{\cML}_{i,j}\cdots d^{\cML}_{M,q}]^{T}\!\in\!\RR^N$, which tracks vectors closest to $\hat{\mbf{c}}^{\GRAND}$, but with bit-flips at corresponding indices. Searching the entire lattice $\bar{\mathcal{X}}$ results in $d^{\ML} \!=\! \min_{\mbf{x} \in \bar{\mathcal{X}}}\norm{\mbf{y} - \mbf{H}\mbf{x}}^{2}$; we can re-express \eqref{eq:LLR_ML2} as
\begin{equation}\label{eq:LLR_ML3}
  \lambda_{i,j}^{\ML} \approx
      \begin{cases}
      \frac{1}{\sigma^{2}} d^{\ML} - \frac{1}{\sigma^{2}} \min_{\mbf{x} \in \bar{\mathcal{X}}^{i,j,0}}{\norm{\mbf{y} - \mbf{H}\mbf{x}}^{2}}  & \text{if $\hat{c}^{\ML}_{i,j}=1$}\\
      \frac{1}{\sigma^{2}} \min_{\mbf{x} \in \bar{\mathcal{X}}^{i,j,1}}{\norm{\mbf{y} - \mbf{H}\mbf{x}}^{2}} - \frac{1}{\sigma^{2}} d^{\ML}  & \text{if $\hat{c}^{\ML}_{i,j}=0$},
    \end{cases} 
\end{equation}
where $d^{\cML}_{i,j} = \min_{\mbf{x} \in \bar{\mathcal{X}}^{i,j,0}}{\norm{\mbf{y} - \mbf{H}\mbf{x}}^{2}}$ if $\hat{c}^{\ML}_{i,j}=1$ and $d^{\cML}_{i,j} = \min_{\mbf{x} \in \bar{\mathcal{X}}^{i,j,1}}{\norm{\mbf{y} - \mbf{H}\mbf{x}}^{2}}$ if $\hat{c}^{\ML}_{i,j}=0$ (note that colored noise scaling is embedded into $\mbf{d}^{\ML}$ and $\mbf{d}^{\cML}$ in Alg. \ref{alg:turboGRAND}). However, turbo-GRAND only searches a limited number of $\bar{\mbf{x}}$ vectors that are extracted from the recovered $\bar{\mbf{c}} \!=\! \hat{\mbf{c}}\ominus \mbf{w}$ words via guesswork, and that can be accumulated in a set $\mathcal{S}$ $\left(\abs{\mathcal{S}} <T\!\times\!B\!\ll\!\abs{\bar{\mathcal{X}}}\right)$.

We first initialize $d^{\ML}$ to $\infty$ and $\mbf{d}^{\cML}$ to saturation noise-scaled thresholds. Then, $d^{\ML}$ and $\mbf{d}^{\cML}$ are updated iteratively, upon every new noise guess (up to $B$ guesses per iteration $t$ in GRAND with abandonment). For every guessed word, $\bar{\mbf{c}}$, we re-generate a modulated vector, $\bar{\mbf{x}} \!=\! \text{mod}\left(\bar{\mbf{c}}\right)$, and add it to $\mathcal{S}$. Hence, for turbo-GRAND, $d^{\ML} \!=\! \min_{\bar{\mbf{x}} \in \mathcal{S}}\norm{\mbf{y} - \mbf{H}\bar{\mbf{x}}}^{2}$, and
\begin{equation}\label{eq:LLR_GRAND}
  \bar{\lambda}_{i,j}^{\GRAND} \!\approx\!
      \begin{cases}
      \frac{1}{\sigma^{2}} d^{\ML} \!-\! \frac{1}{\sigma^{2}} \min_{\bar{\mbf{x}} \in \mathcal{S}^{i,j,0}}{\norm{\mbf{y} \!-\! \mbf{H}\bar{\mbf{x}}}^{2}} & \text{if $\hat{c}^{\GRAND}_{i,j}\!=\!1$}\\
      \frac{1}{\sigma^{2}} \min_{\bar{\mbf{x}} \in \mathcal{S}^{i,j,1}}{\norm{\mbf{y} \!-\! \mbf{H}\bar{\mbf{x}}}^{2}} \!-\! \frac{1}{\sigma^{2}} d^{\ML}  & \text{if $\hat{c}^{\GRAND}_{i,j}\!=\!0$},
    \end{cases} 
\end{equation}
where $d^{\cML}_{i,j} \!=\! \min_{\bar{\mbf{x}}\in \mathcal{S}^{i,j,0}}{\norm{\mbf{y} - \mbf{H}\bar{\mbf{x}}}^{2}}$ if $\hat{c}^{\GRAND}_{i,j}\!=\!1$ and $d^{\cML}_{i,j} \!=\! \min_{\bar{\mbf{x}} \in \mathcal{S}^{i,j,1}}{\norm{\mbf{y} - \mbf{H}\bar{\mbf{x}}}^{2}}$ if $\hat{c}^{\GRAND}_{i,j}\!=\!0$. For the ${i,j}$ indices where $d^{\cML}_{i,j}$ cannot be computed over $\mathcal{S}$, owing to the absence of a corresponding $\bar{\mbf{x}}$, the initial saturated value of $N/\sigma_i^2$ is retained. Note that we use a separate $\bar{d}^{\ML}$ for the decoded $\bar{\mbf{c}}^{\GRAND}$ because not every $\bar{\mbf{x}}$ in $\mathcal{S}$ corresponds to a code-word. Each turbo-GRAND iteration $t$ ends with a single code-word hit $\bar{\mbf{c}}^{\GRAND}(t)$; the final decoded vector after $T$ iterations is
\begin{equation}\label{eq:decoded}
  \bar{\mbf{c}}^{\GRAND} = \argmin_{\bar{\mbf{c}}^{\GRAND}(t); \ t \in \{1,\cdots,T\}} \norm{\mbf{y} - \mbf{H}\ \text{mod}\left(\bar{\mbf{c}}^{\GRAND}(t)\right)}^{2}.
\end{equation}
Note that Algorithm \ref{alg:turboGRAND} does not explicitly construct $\mathcal{S}$ and store it in memory for post-processing but instead updates the ML and counter-ML distance metrics on the fly.

The core of turbo-GRAND is a soft-input decoding mechanism which rank-orders candidate noise sequences according to $\bar{\mbf{\Lambda}}^{\GRAND}$. Let $\mbf{W}\!\in\!\FF_2^{2^N\times N}$ be a matrix containing in its rows all possible noise sequences, and let $\acute{\Pi}\!:\! \RR^{2^N} \rightarrow \{1, \cdots, 2^N\}^{2^N}$ be a sorting function (increasing order). Then, $\mbf{s}=\acute{\Pi}\left(\mbf{W}\times\abs{\mbf{\Lambda}}\right)$ is a vector of sorted noise-sequence indices, and $\mbf{w} = \Pi(\mbf{s}(k))$ retrieves the $\nth{k}$ likely noise sequence. However, populating noise sequences in a single matrix is not hardware-friendly nor computationally efficient. Alternatively, SGRAND \cite{Solomon9149208} recursively constructs a max-heap for each combination of reliabilities in $\mbf{\Lambda}$ to dynamically generate $\mbf{w}$ vectors with increasing likelihoods. Also, ORBGRAND \cite{Duffy9414615} builds a bit permutation map based on the decreasing rank order of bit reliability to generate a pre-determined series of putative noise queries. Most probably, we have $\mbf{s}(i)\!\leq\! \mbf{s}(j)$ when $\Prob(\mbf{w}\!=\!\mbf{w}_i) \!\geq\! \Prob(\mbf{w}\!=\!\mbf{w}_j)$. In SGRAND \cite{Solomon9149208}, the latter is an ``if and only if'' condition. Thus, with either SGRAND or ORBGRAND, turbo-GRAND always queries the all-zeros noise sequence first and is biased to give higher priority to noise sequences of smaller Hamming weight, but not necessarily so, depending on the reliability information in $\bar{\mbf{\Lambda}}^{\GRAND}$. In the absence of soft information or further channel knowledge, noise query follows increasing Hamming weights for a probability of bit flip less than $1/2$. The latter is captured by an SGRAND core of turbo-GRAND, which reduces to hard-GRAND in the absence of soft information. Hence, SGRAND can be used in the first turbo-GRAND iteration when $\mbf{\Lambda} \!=\! \mbf{0}_N$. ORBGRAND requires some soft input information to match the performance of SGRAND, so it can be adopted when $\mbf{\Lambda} \!=\! \mbf{\Lambda}^{\ZF}$ in the first iteration. However, ORBGRAND entails more noise guesses on average, so it has the potential to generate richer LLRs.

Turbo-GRAND can be modified to support iterative disjoint detection and decoding, where $\bar{\mbf{\Lambda}}^{\GRAND}$ is fed all the way back to a separate detector. Towards this end, the detector should itself be SISO. For instance, the modified distance metric of a turbo MAP detector can be \cite{tomasoni2012hardware}
\begin{equation}\label{eq:map}
 \varphi(\mbf{x}) = - \frac{\norm{\mbf{y} - \mbf{H}\mbf{x}}^{2}}{\sigma^{2}} + \sum_{i=1}^{N}\sum_{j=1}^{q}{\left(2\mbf{c_{x}}(i,j)-1\right)\bar{\mbf{\Lambda}}^{\GRAND}(i,j)},
\end{equation}
where $\mbf{c_{x}}$ is the bit representation of $\mbf{x}$. The a posteriori LLRs can be calculated as
\begin{equation}\label{eq:LLR_MAP}
  \lambda_{i,j}^{\MAP} = \left(\max_{\mbf{x} \in \mathcal{X}^{i,j,1}}{\varphi(\mbf{x})} - \max_{\mbf{x} \in \mathcal{X}^{i,j,0}}{\varphi(\mbf{x})} \right).
\end{equation}
Disjoint detection and GRAND offers good modularity, where on every detection iteration, the noise can be filtered towards new signal subspaces \cite{ISIT18_Sarieddeen}, yielding more efficient GRAND. 

\begin{figure*}[ht!]
  \centering
    \subfloat[AWGN - SGRAND turbo-GRAND with nil soft input - BPSK.]{\label{fig:AWGN} \includegraphics[width=0.49\linewidth]{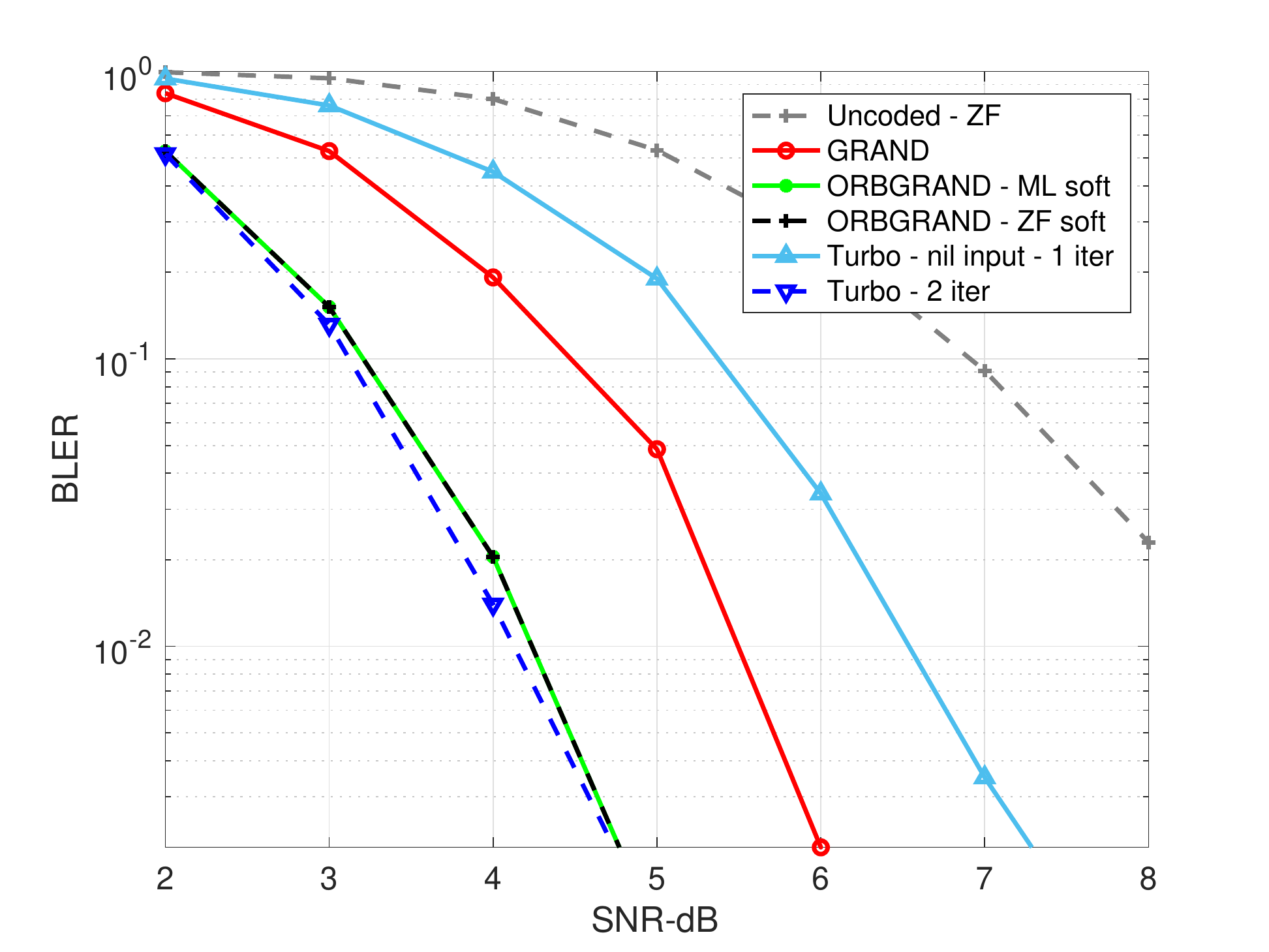}}
  \hfill
  \subfloat[Rayleigh - SGRAND turbo-GRAND with nil soft input - BPSK.]{\label{fig:Rayl} \includegraphics[width=0.49\linewidth]{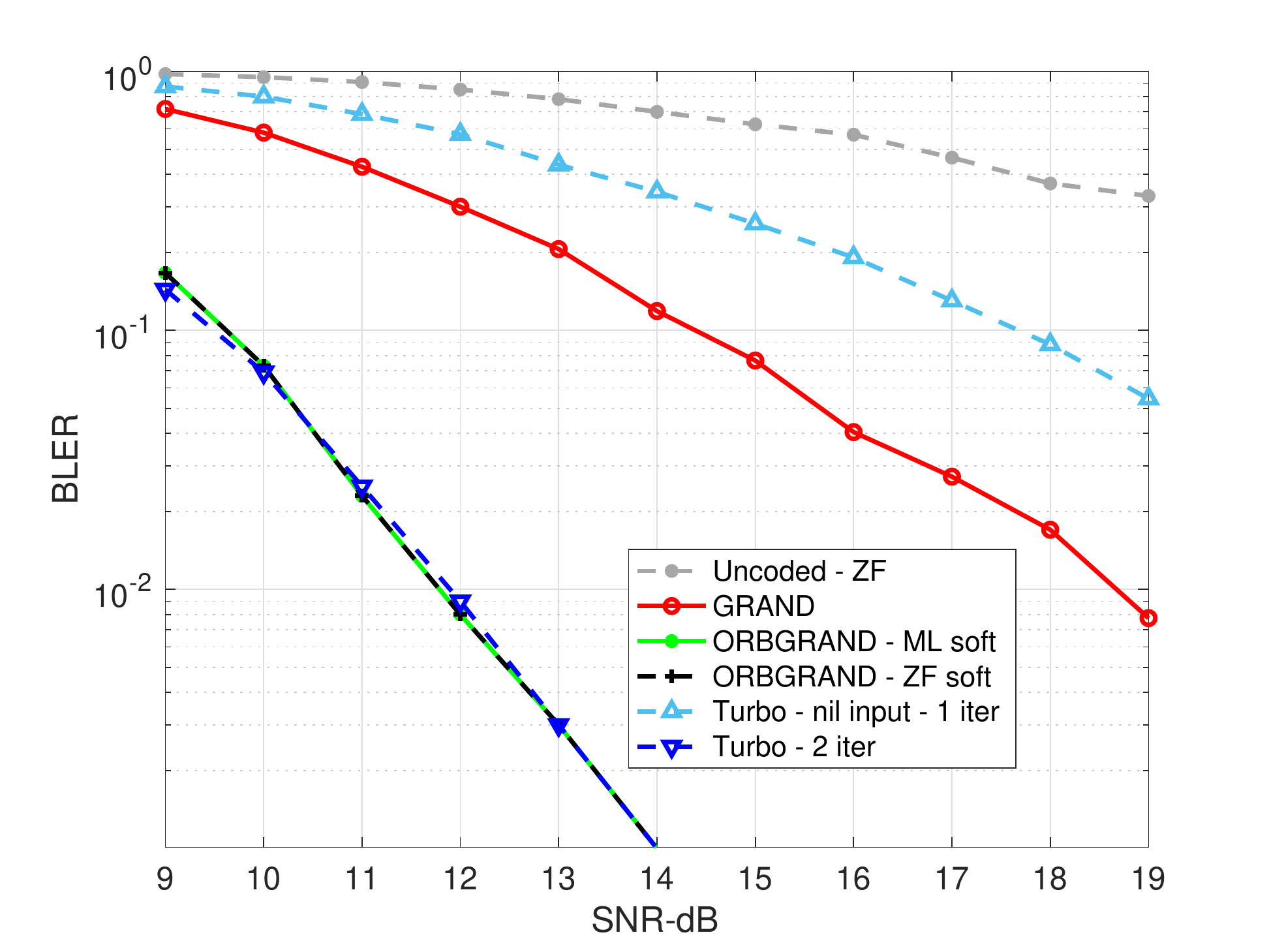}}

  \subfloat[Rayleigh - SGRAND turbo-GRAND with nil soft input - 16QAM.]{\label{fig:16QAM} \includegraphics[width=0.49\linewidth]{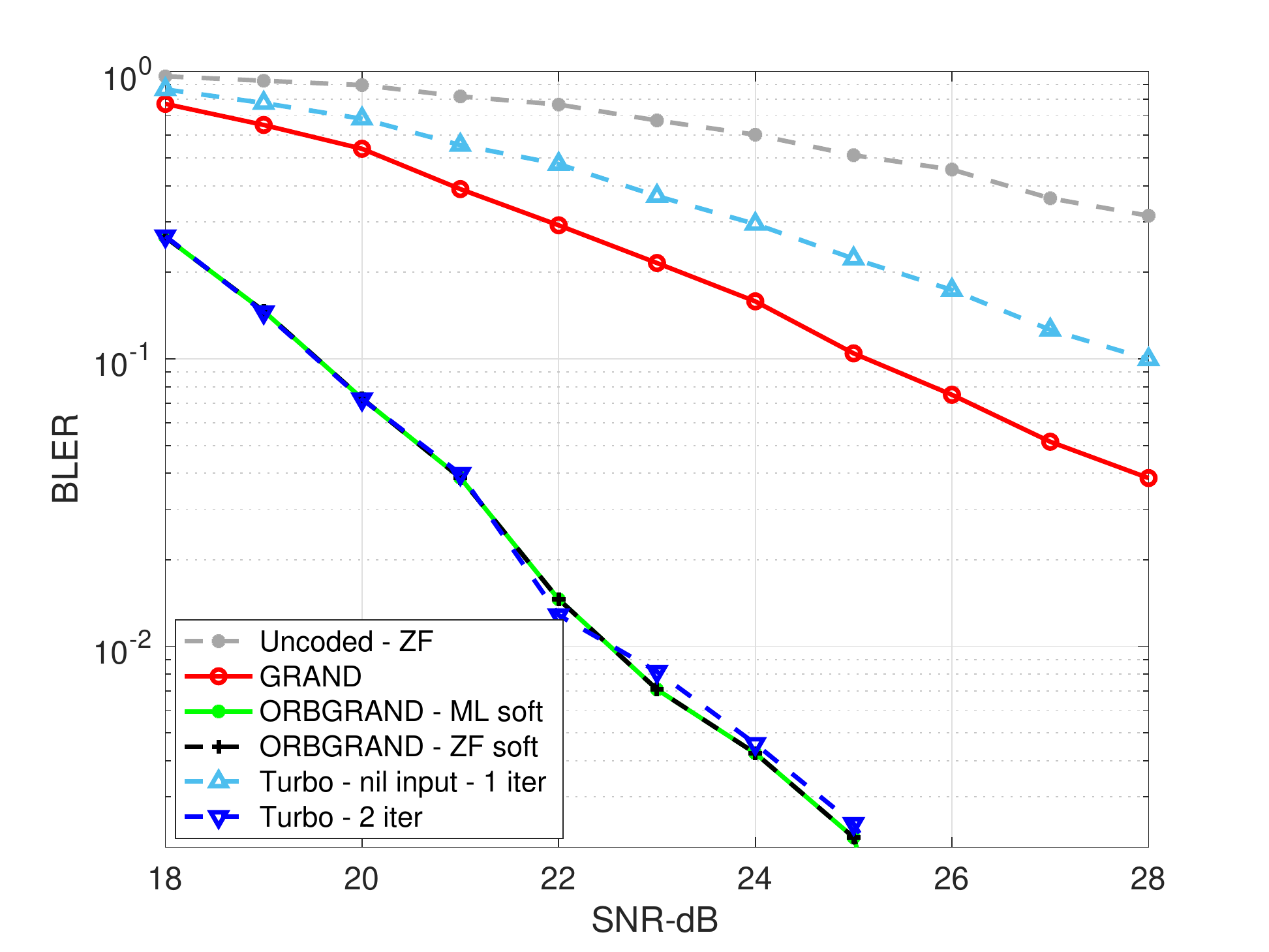}}
  \hfill
  \subfloat[Rayleigh - ORBGRAND turbo-GRAND (ZF soft input) - $90\%$ CSI - BPSK.]{\label{fig:ORB} \includegraphics[width=0.49\linewidth]{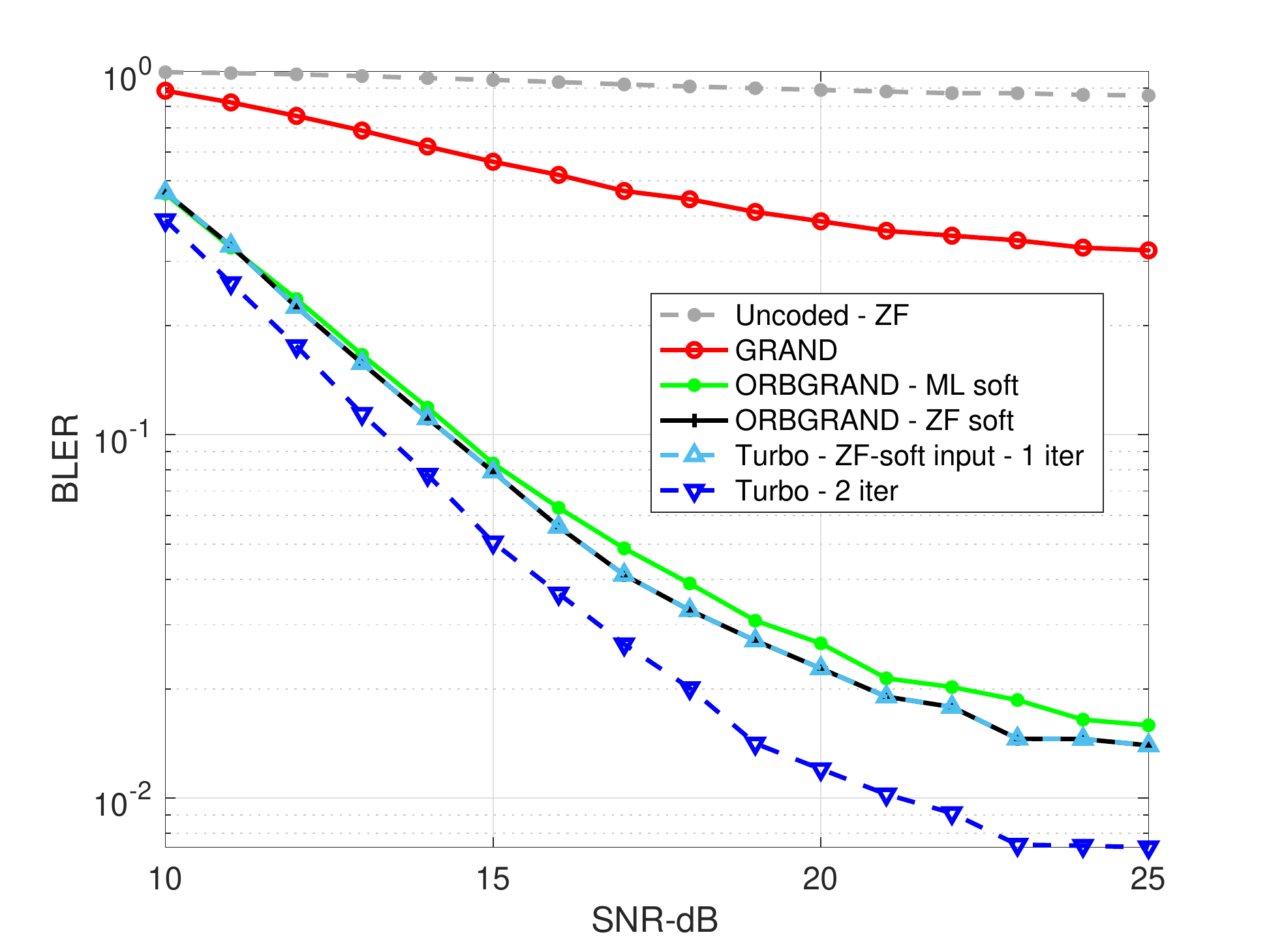}}

  \caption{BLER performance evaluation of the proposed turbo-GRAND schemes with BCH {[\ \!127,113]\ \!} codes.}
  \label{fig:BLER}
\end{figure*}


\section{Performance and Complexity Evaluation}
\label{sec:results}

Following the system model of Sec.~\ref{sec:sysmodel}, the block-error rate (BLER) performance of turbo-GRAND in decoding Bose–Chaudhuri–Hocquenghem (BCH) {[\ \!127,113]\ \!} codes is compared to reference GRAND/ORBGRAND schemes (ORBGRAND exhibited superior performance and complexity tradeoffs  compared to other code-specific decoders \cite{Duffy9414615}). Assuming AWGN and normalized transmit power (SNR of $1/\sigma^{2}$), with the absence of input soft information, Fig. \ref{fig:AWGN} illustrates that SGRAND-based turbo-GRAND generates soft information that matches ML-detection soft information. Turbo-GRAND can outperform ML-detection-based ORBGRAND because: 1) turbo iterations can introduce a list-decoding gain and 2) distance computations are over an entire code-word of many channel uses; feasible ML detection search routines only span a few channel uses. A similar relative performance is noted under Rayleigh fading in Fig. \ref{fig:Rayl} and Fig. \ref{fig:16QAM}, using binary phase-shift keying (BPSK) and 16-quadrature amplitude modulation (16-QAM) (Gray mapping), respectively. The gains in soft-GRAND schemes are larger under fading (more than $\unit[6]{dB}$ SNR gains at a BLER of $10^{-2}$), highlighting the importance of turbo-GRAND in the absence of soft information. We further note that the turbo-GRAND gains are captured in two iterations, beyond which diminishing returns are expected, as both SGRAND and ORBGRAND converge faster on every new iteration, $t$, guessing over an enhanced initial vector, $\hat{\mbf{c}}^{\GRAND}(t)$. The achievable gains of turbo-GRAND can be improved by tuning a fixed guess budget, $B$. Alternatively, some sort of reactive taboo search \cite{Datta_2010} can be adopted, in which every iteration starts with a pseudo-random initial vector upon which noise is guessed.

Several communication system scenarios further highlight the turbo-GRAND gains, especially under symbol interference in spatial/path diversity schemes, where ML soft information is significantly better than ZF soft information. In such scenarios, starting from ZF soft information, ORBGRAND-based turbo-GRAND can bridge the gap to the much more complex ML-detection-based soft-GRAND (this paper only covers uncorrelated point-to-point channels). In another scenario, under imperfect CSI, joint detection and decoding in turbo-GRAND outperforms conventional soft decoding. We assume $10\%$ CSI error in Fig. \ref{fig:ORB}, where $\mbf{H}_{\text{err}} \!=\! 0.9\mbf{H} \!+\! 0.1\tilde{\mbf{H}}$, and $\tilde{\mbf{H}}$ has the same distribution as $\mbf{H}$ but is independently and randomly generated. Starting from ZF soft information as input LLRs, ORBGRAND-based turbo-GRAND outperforms both ML-soft- and ZF-soft-ORBGRAND. Note, however, that our ML-detection implementation in these simulations only undergoes an exhaustive search over subsets of symbols in $\mbf{x}$ of size four.

We next analyze the complexity in terms of floating-point operations in complex multiplication ($\CMT$) and complex addition ($\CAD$). We compare the additional processing in turbo-GRAND over soft-GRAND to the complexity of soft-output ZF and ML detection. The search complexities are dominated by Euclidian distance computations and the complex matrix multiplications they entail. The complexity is exponential (in the symbol-vector length, $M$) with ML detection (\eqref{eq:LLR_ML2} and \eqref{eq:LLR_ML3}), polynomial with turbo-GRAND \eqref{eq:LLR_GRAND}, and linear with ZF \eqref{eq:LLR_ZF}. 

Table \ref{table:complexity} illustrates the approximate worst-case complexity of generating soft information in one channel use. The average complexity of turbo-GRAND is much less because the guess budget, $B$, is not exploited on every iteration; with more iteration and higher SNR, a few or a single guess can recover a code-word. Turbo-GRAND is thus much less complex than conventional iterative list-based detection and decoding. Even with larger guess budgets, the recovered words on different iterations often overlap, and redundant computations can be saved. Furthermore, because noise-sequence guessing typically follows increasing Hamming weights, the vector Euclidean distance computations can reduce to simple symbol-based scalar distance computations/updates, resulting in an average linear turbo-GRAND complexity of $T\times B\left(M\CMT + M\CAD\right)$. Hence, for large modulation orders, a hardware-optimized turbo-GRAND can even prove to be less complex than soft-output ZF detection followed by soft-GRAND; much less complex than ML detection. Further simplifications can be made if the channel remains static over multiple uses.


\begin{table}[!t]
\caption{Soft-output data detection complexity comparison} 
\label{table:complexity} 
\centering 
\begin{tabular}{|c|| c | }
  \hline
  Detector & Complexity \\
  \hline\hline
  ML &  $\abs{\mathcal{X}}^M\left((M^2\!+\!M)\CMT + (M^2\!+\!M)\CAD\right)$ \\
  \hline
  ZF &  $\abs{\mathcal{X}}\left(M\CMT + M\CAD\right)$ \\
  \hline
  turbo-GRAND &  $T\times B\left((M^2\!+\!M)\CMT + (M^2\!+\!M)\CAD\right)$ \\
  \hline
\end{tabular}
\end{table}


\section{Conclusions}
\label{sec:conc}

We proposed a mechanism for using GRAND to extract soft information in a joint detection and decoding framework. By leveraging access to complex received symbols, hard demapped bits, CSI, and possibly noise statistics, we generate LLRs by populating a shortlist of Euclidean distance computations. Such LLRs can be used in subsequent soft-decoding GRAND iterations, giving rise to turbo-GRAND. Compared to hard GRAND, a few iterations of turbo-GRAND introduce an excess of $\unit[6]{dB}$ SNR gains at a BLER of $10^{-2}$ (much higher gains at lower BLERs), under practical communication system scenarios of Rayleigh fading channels. Furthermore, turbo-GRAND can match and even outperform exhaustive ML-detection-based soft-GRAND at a much-reduced average linear complexity. This work can extend into a generic joint detection and decoding framework for future investigations.



\end{document}